\documentclass[]{spie}  

\usepackage{ wasysym }
\usepackage{amsfonts,amssymb}
\usepackage{graphicx}
\usepackage[colorlinks=true, allcolors=blue]{hyperref}
 \usepackage[table,xcdraw]{xcolor}
\title{A Design Study on Adaptive Primaries for 1-2 Meter Class Telescopes}

\author[a]{J. Fowler}
\author[a]{Rachel Bowens-Rubin}
\author[a]{Philip M. Hinz}
\affil[a]{University of California Santa Cruz, 1156 High Street, Santa Cruz, United States}

\authorinfo{Further author information: (Send correspondence to J. Fowler)\\J. Fowler: E-mail: jumfowle@ucsc.edu, Telephone: 1 512 963 9951\\ }

\begin{document} 
\maketitle

\begin{abstract}
Adaptive optics (AO) offers an opportunity to stabilize an image and maximize the spatial resolution achievable by ground based telescopes by removing the distortions due to the atmosphere.  Typically, the deformable mirror in an AO system is integrated into the optical path between the secondary mirror and science instrument; in some cases, the deformable mirror is integrated into the telescope itself as an adaptive secondary mirror. However including the deformable mirror as the primary mirror of the telescope has been left largely unexplored due to the previous cost and complexity of large-format deformable mirror technology. In recent years this technology has improved, leaving deformable primary mirrors as a viable avenue towards higher actuator density and a simplification in testing and deploying adaptive optics systems. We present a case study to explore the benefits and trade-offs of integrating an adaptive optics system using the primary mirror of the telescope in small-to-mid-sized telescopes.  
\end{abstract}


\section{INTRODUCTION}
\label{sec:intro}  


\subsection{The State of Adaptive Primaries for Astronomical Observations}

As of the writing of this paper, no ground-based telescopes used for astronomical observations are built with an adaptive primary mirror.  The first discussion of an adaptive primary mirror in the literature appears in Brusa 1999, ``From adaptive secondary mirrors to extra-thin extra-large adaptive primary mirrors.''~\cite{Brusa1999}  This paper, along with Riccardi 2003 et al.~\cite{Riccardi2003}, lay the foundations for converting adaptive secondary mirror technology for use as an adaptive primary.  Riccardi 2004~\cite{Riccardi2004} explores the idea of building adaptive primary segments into  ELT class telescopes in order to obtain a high-density actuator AO system, but ultimately, this path was not pursued in any final designs.      
These concept studies were written assuming voice-coil actuator adaptive secondary mirror technology, which is limited by thin mirror facesheets and the lower power generation of those actuators. 

Active optics systems, which hold the optical prescription of the large mirrors under changing gravitational and environmental conditions, are in place at several major observatories. ``Active optics'' systems are distinct from ``adaptive optics'' systems because they cannot apply corrections quickly enough to correct the atmospheric turbulence evolution in real time. 
Designs for the European Extremely Large Telescope (E-ELT) include considerations for active optics\cite{Messenger2007}. Similarly, active optics are in use for the New Technology Telescope (NTT)\cite{Messenger53} and the Nordic Optical Telescope (NOT)\cite{NOT1992}. 
Finally, Keck Observatory can move its primary mirror segments to maintain a phased optical system (applying piston, tip, and tilt)\cite{Chanan2000}, though they are moved only for calibration and engineering testing and not available to update during typical science observations. 




\subsection{Future Adaptive Secondary Mirrors Constructed with Hybrid-Reluctance Variable Actuators}

The Netherlands Organization for Applied Scientific Research (TNO) has made significant breakthroughs simplifying large-format deformable mirror technology. 
The key advancement is a new style of hybrid variable reluctance (HVR) actuator that is $\sim$75 times more efficient than traditional voice-coil actuators (HVR actuator efficiency~\cite{Kuiper2018} = $38 N/\sqrt{W}$; for comparison the Multiple Mirror Telescope (MMT) and Large Binocular Telescope (LBT) have an actuator efficiency of ~\cite{Riccardi2003} = $0.5 N/\sqrt{W}$). The lower required power offers a pathway to increase the facesheet thickness and operate large-format deformable mirrors without dedicated cooling systems, allowing the technology to be applied  in a greater diversity of situations. 

\begin{figure}[h]
   \begin{center}
   \begin{tabular}{c}
   \includegraphics[width=0.6\textwidth]{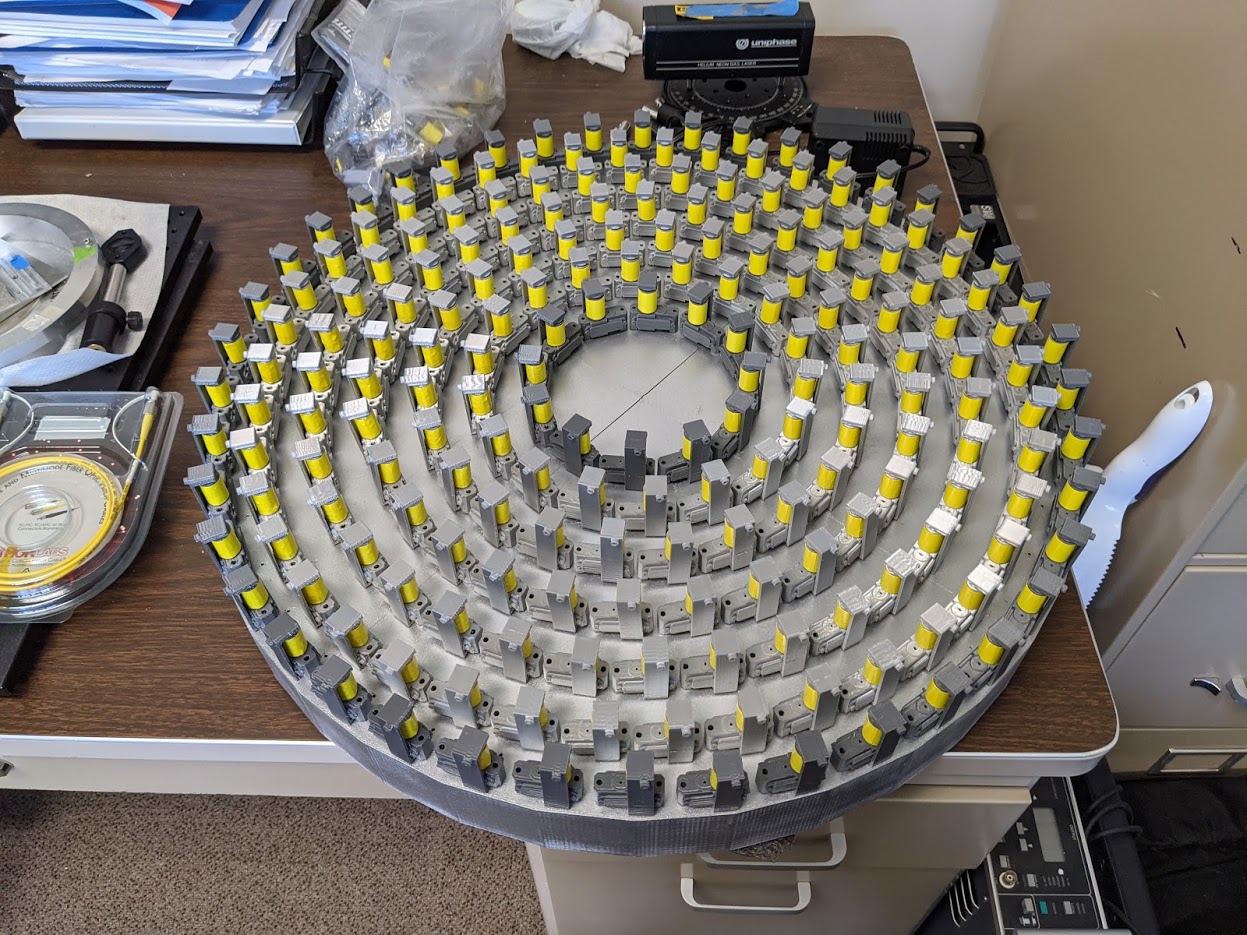}
   \end{tabular}
   \end{center}
   \caption[example] 
   { \label{fig:UH883Dprint} 
   \textbf{3D Printed Replica of the UH2.2m Adaptive Secondary Mirror}. The construction of the UH2.2m adaptive secondary mirror will be completed in 2022. The UH2.2m ASM will demonstrate the ability to use large-format deformable mirrors constructed with the HVR actuator technology to complete adaptive optics corrections in mid-sized telescopes. }
   \end{figure}

The first on-sky adaptive secondary mirror (ASM) constructed with HVR actuators will be deployed to the University of Hawaii 2.2-meter telescope (UH2.2m) on Mauna Kea, HI\cite{Chun2020}.   The assembly and integration is currently underway for the UH2.2m ASM, and it will deploy in 2022. The UH2.2m ASM contains 210 actuators within a secondary  diameter of 62cm (Figure \ref{fig:UH883Dprint}). This corresponds to a physical actuator spacing of 39mm and a projected primary actuator spacing (i.e., the spacing of the actuators as they would be reflected on the primary mirror) of 138mm.

Additionally, an adaptive secondary mirror is under consideration for the W.M. Keck Observatory using HVR actuator technology.  An initial concept study for the Keck ASM was funded in August 2020, and a Phase A study was funded in August of 2021.  The design considerations for the Keck ASM are between 1000-4000 actuators within a diameter of 1.4m, The design with an actuator count of 2087 is currently the most favored\cite{Hinz2020}.   This corresponds to a physical actuator spacing of 26mm and a projected primary spacing of 245mm.  
   
 \subsection{Building an Adaptive Primary Mirror}
 
 Optics used for telescope mirrors are typically fabricated via grinding and polishing flat, low-expansion glass blanks into a static, precision curved surface. This process requires significant time and labor. 
 As demonstrated with the successes of Active Optics on NNT and NOT, the ability to offload some of the optical precision to actuators on the primary would relax optical tolerances of the primary, and drive the cost of the instrument lower. 
 
 As an alternative to grinding processes, multiple institutions are exploring glass-slumping techniques to create mirror facesheets for large-format deformable mirrors.  These facesheets are thin ($\sim$3mm), and when assembled into the deformable mirror, can be moved by a set of actuators into an optical precision roughly two orders of magnitude closer to the desired optical shape.~\cite{BowensRubin2020}.  
The facesheets are made via a hot forming technique using inexpensive flat borofloat plates as the starting point for the optic.  Steel ring moulds and top weights are used to generate the final required conic surface. The Lab for Adaptive Optics at UC Santa Cruz is exploring using glass-slumping manufacturing on facesheets up to 1.5m in diameter\cite{Hinz2020} and expect that the approach could easily be adapted for a primary mirror facesheet up to 2m.  
Combined with suitable actuators, this process outlines a straightforward approach to an adaptive primary telescope system, replacing the traditional static primary for comparable costs.

\section{BASIS FOR OUR MODELLING}

Typical models of adaptive optics (AO) systems project the actuator count and spacing to the primary mirror and simulate the performance as if the actuators deform the primary mirror of this system. For this reason, we perform comparable simulations and performance estimates, simply varying the effectiveness of the actuators and their spacing if no projection was needed. Appendix \ref{sec:ao_appendix} outlines the mathematical basis for our AO error budget calculations and simulations. 

\subsection{Baseline for Performance Simulations}

As a starting point for our simulations, we use the site conditions of Mauna Kea, home to the Keck Observatory and UH2.2m. 
Taking inspiration from the proposed Keck ASM and UH2.2m ASM makes setting our simulations at Mauan Kea a logical comparison. The site parameters for Mauna Kea, as well as the other defaults for our simulations (including the default speed of the AO control loop and our science wavelength) are outlined in Table \ref{tab:ao_default_params}. For more detailed descriptions of each parameter see Appendix \ref{sec:ao_appendix}. Our analysis for this study is performed with a new software package \texttt{WIGGLE}, Wavefront Improvements with GiGantic Light-deforming Experiments, which is described in Appendix \ref{sec:wiggle_appendix}.

   \begin{table}[ht]
\caption{\textbf{Static AO Parameters for the Performance  Simulations.} 
} 
 \label{tab:ao_default_params}
 \begin{center}       
 \begin{tabular}{|l|l|l|l|}
 \hline
 \rule[-1ex]{0pt}{3.5ex}Parameter & Definition & Value & Units \\
\hline\hline
telescope diameter & primary mirror diameter & 1.4 & meters \\
\hline
science wavelength & wavelength at which science is conducted & 2500 & nanometers \\
\hline
coherence cell size  & physical resolution of turbulence (r$_0$) & 15 & centimeters \\
\hline
zenith angle & angle at which science object is observed & 20 & degrees \\
\hline
wind speed & speed of the wind on site & 10 & meters/second \\
\hline
science object separation & distance between the guidestar and the science object & 0 & degrees \\
\hline
readnoise & noise during camera readout & 1 & electrons/pixel \\
\hline
guide star magnitude & apparent magnitude of the star used for guiding & 8 & V band mag \\
\hline
controller frequency & speed at which AO calculation can run & 50 & Hertz \\
 \hline 
 \end{tabular}
 \end{center}
 \end{table}
 
\section{ENGINEERING DISCUSSION AND SIMULATIONS}
\label{sec:sections}

\subsection{Outline of Analysis}

We start our analysis by looking at the technology under development by The Netherlands Organization of Applied Science (TNO) for the Keck adaptive secondary mirror. At present, a 1.4 meter optic with 2087 actuators and a 26mm spacing\cite{Hinz2020} is a feasible design by engineering standards (i.e., the actuators can support and drive the mirror effectively). Table \ref{tab:keck_asm_specs} shows the AO residuals and engineering parameters for a system in which we simulate the most-favored Keck ASM concept as an adaptive primary. 

   \begin{table}[ht]
\caption{\textbf{AO Error Residuals and Engineering Terms for Keck ASM concept}} 
 \label{tab:keck_asm_specs}
 \begin{center}       
 \begin{tabular}{|l|l|l|l|l|l|l|}
 \hline
 \rule[-1ex]{0pt}{3.5ex}Primary & Actuator &  Facesheet & Quilting & Resonant & High order & Strehl (2.5 $\mu$m) \\
 diameter & spacing & thickness & error & frequency & wavefront error & (fractional  \\
 (m) & (cm) & (mm) & (nm) & (Hz) & (nm) & intensity) \\
\hline\hline
1.4 & 2.6 & 2.9 & 0.26 & 27128.9 & 194.7 & 0.79 \\ 
 \hline 
 \end{tabular}
 \end{center}
 \end{table}

\begin{figure}[h!]
   \begin{center}
   \begin{tabular}{c}
   \includegraphics[width=0.5\textwidth]{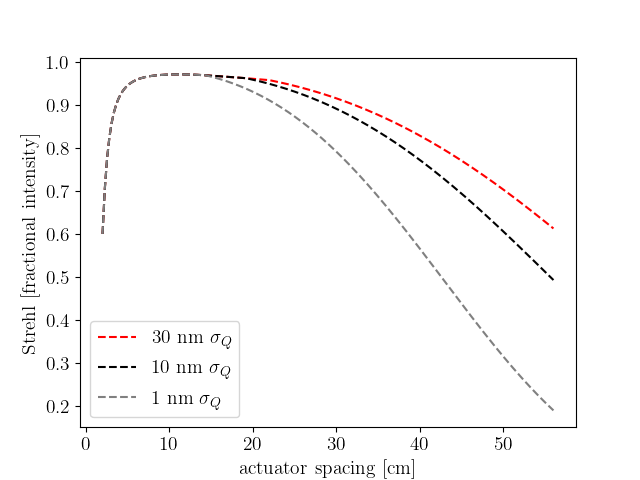}   \includegraphics[width=0.5\textwidth]{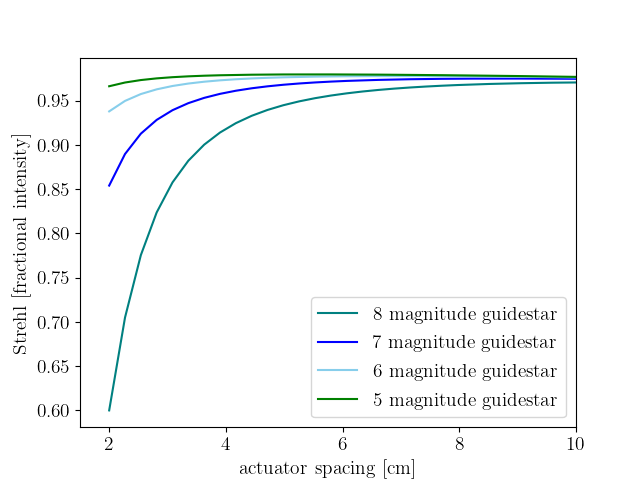}
   \end{tabular}
   \end{center}
   \caption[example] 
   { \label{fig:AO_strehl} 
   \textbf{Changes in Strehl with actuator spacing.} Left: Strehl at a 2500nm and wavelength for 1nm, 10nm, and 30nm of quilting error as actuator spacing increases for an 8 magntiude guidestar. As actuator spacing increases, the Strehl decreases minimally until the resonant frequency of the actuators impact performance. Minimizing the number of actuators only accounts for $\sim0.5\%$ loss of Strehl up to an actuator spacing of 20cm. Right: Strehl at performance with actuator spacing at varying guide star magnitudes and 30nm of quilting error. As the guidestar gets fainter, wavefront sensors have less photons and therefore information to appropriately correct the AO loop for tightly packed actuators. A more complex AO system could address this issue with different modes for different guide stars.}
   \end{figure}

In order to correct Mauna Kea-like seeing conditions (with turbulence on the order of $r_0 = 15$cm) with a mid-sized telescope, an adaptive primary mirror will need fewer actuators than the Keck adaptive secondary mirror design ($<$2087 actuators). 
Having an actuator spacing comparable to the turbulence size will adequately control the atmosphere without over-engineering the number of actuators.\cite{Hardy1998} Furthermore, for fainter stars, measurement error could be impacful at tighter actuator spacings. Figure \ref{fig:AO_strehl} demonstrates this, showing that there is minimal Strehl (fractional intensity of light recovered from an adaptive optics observation) improvement starting with an actuator spacing of 15cm and stepping up to more tightly packed actuators. For our study we focus on a Strehl calculated at 2500nm, as an infrared wavelength comparable to the Keck K-band. 
Similarly, with an increased actuator count comes increased complexity in characterizing the deformable mirror and increased costs to manufacture. Therefore, we take an unusual approach of minimizing actuator count with the rest of this analysis. We examine where the AO residuals and science performance are not degraded while 
maintaining feasible 
engineering designs. 

As we increase the spacing between actuators on the adaptive primary mirror, our system is at risk of increased error from gravitational quilting\cite{Riccardi2003}; the mirror may experience pockets between actuator pokes like panels of a quilt. Equation \ref{eq:quiltling} describes how we calculate quilting error ($ \sigma_{Q}$) for a given mirror design with actuator spacing $d_a$, mirror facesheet thickness $t$, and resistance coefficient of glass $K$:  
\begin{equation} \label{eq:quiltling}
    \sigma_{Q} = \frac{(0.015876)  {d_a}^4}{K t^2}
\end{equation}

\noindent For our calculations we use a $K = 3.24*10^6$m for a borosilcate facesheet material \cite{Riccardi2003}.

We fix the actuator spacing for a given design and its simulation, so quilting error decreases with increasing mirror thickness; a thicker and stiffer mirror will better hold its shape despite more space between actuators. From Equation \ref{eq:quiltling} we estimate the Keck ASM prototype would have $\leq 1$ nm of quilting error. Through the rest of our study, we examine the mirror thicknesses set by allowing a quilting error of 1nm, 10nm, and 30nm to understand how the facesheet thickness impacts the ability of the actuators to drive the mirror.

With a set mirror thickness and actuator spacing, we estimate the mirror mass each actuator must drive, as given by Equation \ref{eq:resonant_freq}, with borofloat density $\rho=2.33$ g/cm$^3$, mirror facesheet thickness $t$, and actuator spacing $d_a$. In Equation \ref{eq:resonant_freq} we divide the the mirror into square sections each driven by a single actuator (i.e., subapertures). Given TNO's HVR actuators, each actuator is driven with a known spring constant and therefore can be modeled as a simple harmonic oscillator; each actuator and mass combination has a resonant frequency. As the mirror thickness increases, the mass that each actuator drives increases, the resonant frequency of the actuator decreases, and reduces the speed at which the AO system can be effectively controlled.  Equation \ref{eq:resonant_freq} describes how we calculate the resonant frequency for our system, with resonant frequency $f_r$, driven mass $M_a$, and actuator spring constant $\kappa$:

\begin{equation}\label{eq:resonant_freq}
f_r = \sqrt{k/M_a} = \sqrt{\kappa /(\rho t {d_a}^2)}
\end{equation}

\noindent We calculate the spring constant for the TNO HVR actuator with
$\kappa = \frac{\mathrm{force}}{\mathrm{displacement}}$. TNO's recorded force is 8N over the linear range, but with a force amplification of 2.6 from an internal lever arm that amounts to 21N\cite{Hinz2020}, for a final $\kappa= 525000$N/m. As the control frequency nears the resonant frequency of the actuators, it risks introducing noise into the control loop, as described in ``Digital Control of Dynamic Systems.''\cite{Franklin1997}
We assume that when the resonant frequency is less than a factor of 10 greater than the control frequency it begins to limit the ability to run the AO loop.

In our trade study, 
we step through the actuator spacing favored by the Keck ASM concept as a primary mirror and move to an actuator spacing that is more effective (i.e. accurately controls $r_0=15$cm regime without notable loss of Strehl). We examine where the AO performance begins to degrade due to what engineering factors and optimize an ideal design for an adaptive primary. 

\subsection{Mirror Thickness with Increased Actuator Spacing}
\label{sec:eng_outline}

\begin{figure}[h!]
   \begin{center}
   \begin{tabular}{c}
   \includegraphics[width=0.5\textwidth]{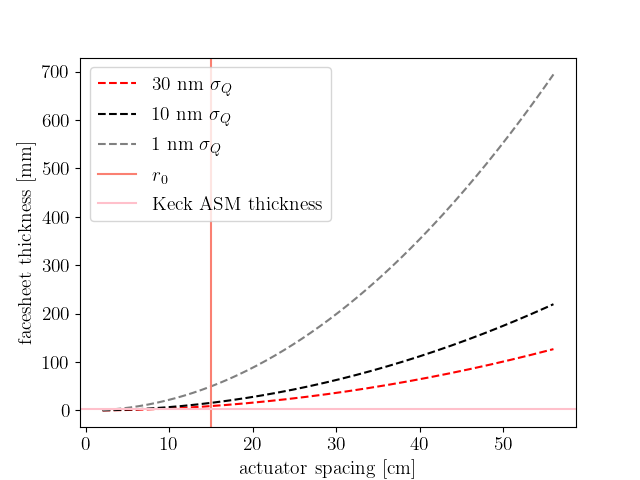}
     \includegraphics[width=0.5\textwidth]{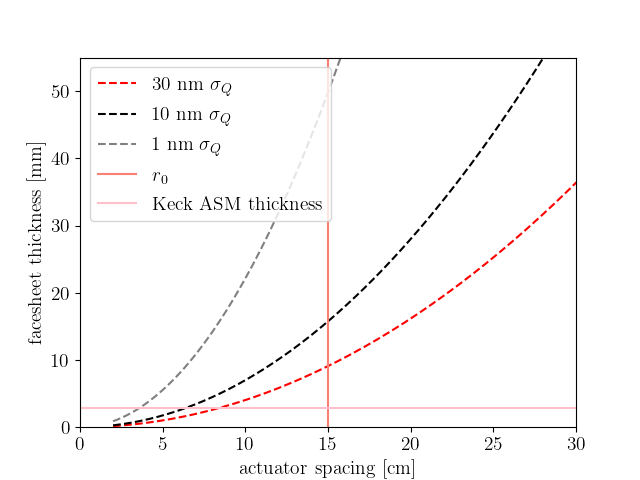}
   \end{tabular}
   \end{center}
   \caption[example] 
   { \label{fig:thickness_v_spacing} 
   \textbf{Minimum mirror thickness for quilting errors of 1nm, 10nm, and 30nm.} 
   At an actuator spacing of 15cm, we require a thickness of 50.0mm at $\sigma_Q = 1$nm, 15.8mm at $\sigma_Q = 10$nm, and 9.1mm at $\sigma_Q = 30$nm. If the quilting error requirement is relaxed to $\sigma_Q =100$nm, the required thickness is 5.0mm, which is comparable to the shell thickness of the Keck ASM concept (at $\sim$3mm.) The thickness of the Keck ASM concept (at 2.9mm) and the turbulence scale of the system ($r_0=15$cm are plotted for reference with the pink and salmon lines.) }
   \end{figure}

For our study we examine three regimes of allowable quilting error: (1) $30$nm, (2) $10$nm, and (3) $1$nm. 
Figure \ref{fig:thickness_v_spacing} shows the required mirror thickness as actuator spacing increases for all three quilting error regimes. Mirror facesheet thickness is proportional to the square of the actuator spacing ($t \propto {d_a}^2$). Figure \ref{fig:thickness_v_spacing} 
demonstrates this relationship between required facesheet thickness as actuator spacing increases. The point where the actuator spacing is comparable to turbulence scale ($d_a = r_0 = 15$cm) intersects with required thicknesses that are approachable facesheet designs. We believe the UCSC Lab for Adaptive Optics could create a $50.0$mm thick facesheet\cite{Hinz2020} (the thickness for the 1nm quilting error condition at an actuator spacing of 15cm). Furthermore, at a more relaxed value of gravitational quilting error $\sigma_Q=100$nm, the required thickness ($\sim5$mm) approaches nominal designs for current adaptive mirror technology.

\subsection{Optimal Controllable Frequency with Increased Actuator Spacing}

\begin{figure}[h!]
   \begin{center}
   \begin{tabular}{c}
   \includegraphics[width=0.7\textwidth]{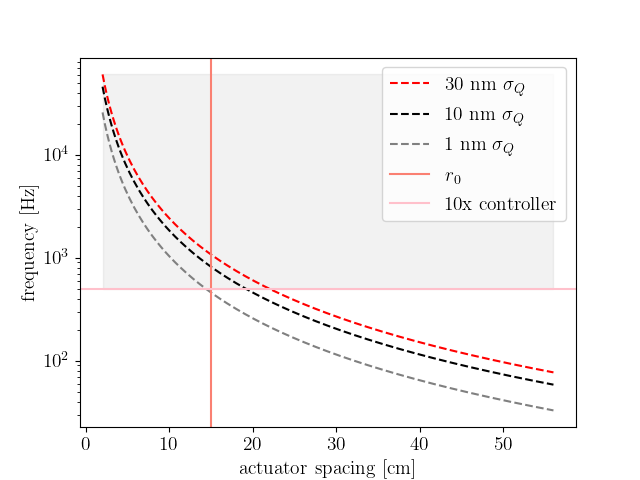} 
   \end{tabular}
   \end{center}
   \caption[example] 
   { \label{fig:freq_v_spacing} 
   \textbf{Resonant frequency with increasing actuator  spacing.} The 10nm and 30nm quilting error regimes have resonant frequencies higher than 10 times the optimal control frequency (500Hz in pink) for an actuator spacing equal to $r_0$ (15cm in salmon). The gray region indicates the resonant frequencies (above 10 times the proposed control frequency of 50Hz) that will not limit the speed of the AO loop.  
  }
   \end{figure}

As the actuator spacing increases with a constant value for quilting error, the resonant frequency of the actuators driving the mirror segment decreases 
(Equation \ref{eq:resonant_freq}).  
Figure \ref{fig:freq_v_spacing} shows this relationship for the 30nm, 10nm, and 1nm error regimes. While resonant frequency impacts adaptive optics performance (as discussed in the next subsection), a 15cm actuator spacing is feasible before the control frequency is impacted for the 10nm and 30nm quilting error regimes.  

\subsection{Control Frequency as a Limiting Factor in AO Performance}

\begin{figure}[h!]
   \begin{center}
   \begin{tabular}{c}
   \includegraphics[width=0.51\textwidth]{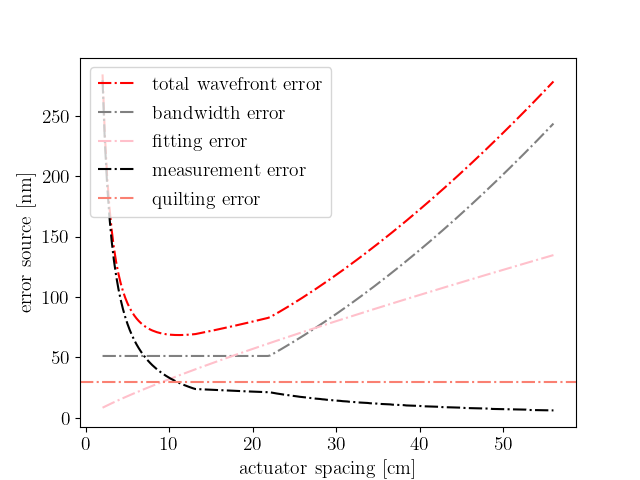}
     \includegraphics[width=0.51\textwidth]{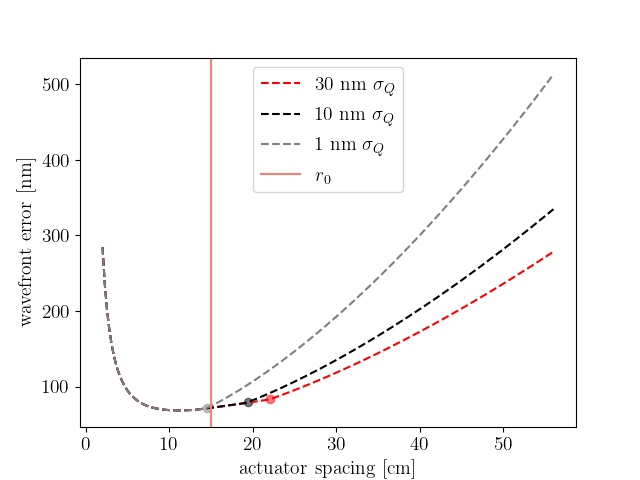}
   \end{tabular}
   \end{center}
   \caption[example] 
   { \label{fig:all_error} 
   \textbf{Sources of error with increasing actuator spacing.} Left: Bandwidth, fitting, and measurement error with spacing as they compare with the total wavefront error in the system. Bandwidth error (in these plots calculated with 30nm allowed quilting error) is the most impactful term in the determination of high order wavefront error, and therefore Strehl. Right: Total wavefront error for the 30nm, 10nm, and 1nm quilting error regimes. The circles indicate the turning point where the resonant frequency begins to limit the AO control loop. At this point the wavefront error increases more significantly with actuator spacing.}
   \end{figure}

To calculate the high order wavefront error as it impacts the total Strehl, we add the following AO error terms in quadrature: 
fitting error (how well we can apply a correction to a wavefront), measurement error (how well we can observe the shape of a wavefront), and bandwidth error (how well we can drive the controller before the wavefront changes). See Appdendix \ref{sec:ao_appendix} for further details. As shown in Figure \ref{fig:all_error}, we find that bandwidth error is 
the dominant term in the total wavefront error calculation when the system is impacted by the resonant-frequency limiting rate. 

\subsection{Optimized Designs and Feasibility of an Adaptive Primary for Mid-sized Telescopes}

\begin{figure}[h!]
   \begin{center}
   \begin{tabular}{c}
   \includegraphics[width=0.51\textwidth]{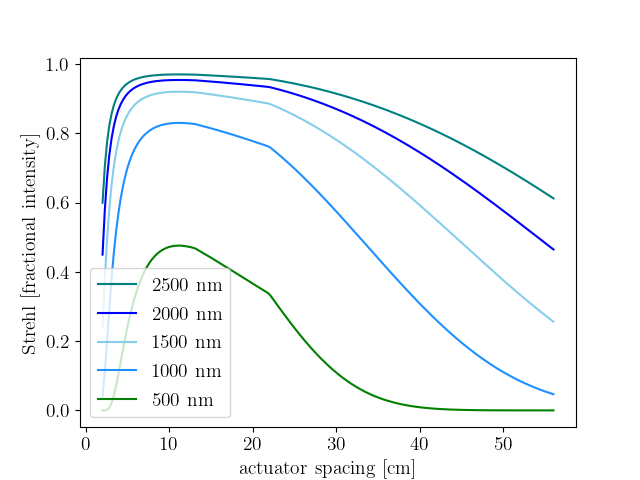}   \includegraphics[width=0.51\textwidth]{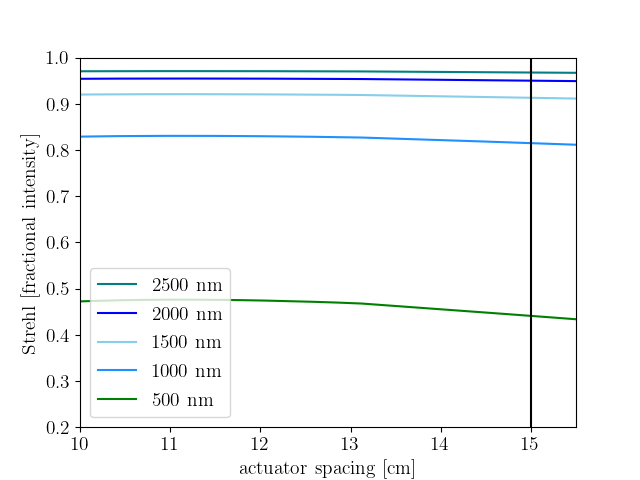}
   \end{tabular}
   \end{center}
   \caption[example] 
   { \label{fig:strehls_wvs} 
   \textbf{Strehl performance at varying science wavelengths} with quilting error of 30 nm. For this analysis we focused on infrared Keck-like observations set at a science wavelength of 2500nm ($\sim$K-band). However, minimizing actuators at other science wavelengths does not notably impact performance over the length scale of the turbulence. Even at a science wavelength of 500nm, we see only $\sim5\%$ Strehl decrease over the region of our final optimized design. }
   \end{figure} 
   
  The premise of this design study began with a technically sound engineering concept: the Keck adaptive secondary mirror concept can be created with existing technology using TNO's HVR actuators and could operate as an adaptive primary mirror.  
When adapting the Keck ASM design ($\diameter$1.4m, 2087 actuators) to a mid-sized adaptive primary design, we find that having as many as 2087 actuators would be unnecessary and the number of actuators could be reduced to save costs without impacting AO performance in the infrared. 
The performance of the AO controller (from the perspective of 
reducing wavefront error) is not significantly reduced until you reach the regime of actuator spacing on the order of $r_0=15$cm.
When decreasing the number of actuators towards this spacing, quilting error may be a risk if the facesheet thickness is not increased to mitigate it.  When the facesheet thickness is increased, there will be an impact on the speed at which we can operate an AO loop. 
We find that using an actuator spacing 
in the regime of $r_0=15$cm is feasible to engineer without risking major contributions from quilting error or bandwidth error.

\begin{table}[h!]
\caption{\textbf{AO Error Residuals and Engineering Terms for Designs of Note.} The blue rows indicate optimal designs, and red rows indicate where the resonant frequency limits the controller frequency for the AO loop. All designs assume a primary mirror diameter of 1.4m. } 
 \label{tab:final_ao_calc}
 \begin{center}    
\begin{tabular}{|l|l|l|l|l|l|}
 \hline
 \rule[-1ex]{0pt}{3.5ex}Actuator &  Facesheet & Quilting & Resonant & High order & Strehl (2.5 $\mu$m) \\
 spacing & thickness & error & frequency & wavefront error & (fractional  \\
 (cm) & (mm) & (nm) & (Hz) & (nm) & intensity) \\
 \hline\hline
2.6   & 1.5    & 1.0  & 15255.68 & 194.7  & 0.79 \\
\hline
5.41  & 6.48   & 1.0  & 3522.89  & 89.0   & 0.95 \\
\hline
\rowcolor[HTML]{B5D9DA} 
8.22  & 14.96  & 1.0  & 1525.89  & 71.62  & 0.97 \\
\hline
\rowcolor[HTML]{B5D9DA} 
11.03 & 26.94  & 1.0  & 847.43   & 68.56  & 0.97 \\
\hline
\rowcolor[HTML]{B5D9DA} 
13.84 & 42.41  & 1.0  & 538.24   & 70.36  & 0.97 \\
\rowcolor[HTML]{B5D9DA} 
13.84 & 13.41 & 10.0 & 957.14 & 70.36 & 0.97 \\
\rowcolor[HTML]{B5D9DA} 
13.84 & 7.74 & 30.0 & 1259.66 & 70.36 & 0.97 \\
\hline
\rowcolor[HTML]{E4D3D2} 
16.65 & 61.39  & 1.0  & 371.89   & 84.19  & 0.96 \\
\rowcolor[HTML]{B5D9DA} 
16.65 & 19.41  & 10.0 & 661.32   & 74.47  & 0.97 \\
\rowcolor[HTML]{B5D9DA} 
16.65 & 11.21 & 30.0 & 870.35 & 74.47 & 0.97 \\
\hline
\rowcolor[HTML]{B5D9DA} 
19.46 & 83.85  & 1.0  & 272.24   & 103.01 & 0.94 \\
\rowcolor[HTML]{E4D3D2} 
19.46 & 26.52  & 10.0 & 484.12   & 79.71  & 0.96 \\
\rowcolor[HTML]{E4D3D2} 
19.46 & 15.31  & 30.0 & 637.14   & 78.9   & 0.96 \\
\hline
\rowcolor[HTML]{E4D3D2} 
22.27 & 109.82 & 1.0  & 207.87   & 124.21 & 0.91 \\
\rowcolor[HTML]{E4D3D2} 
22.27 & 34.73  & 10.0 & 369.65   & 92.63  & 0.95 \\
\rowcolor[HTML]{E4D3D2} 
22.27 & 20.05  & 30.0 & 486.49   & 84.2   & 0.96 \\
\hline
\rowcolor[HTML]{E4D3D2} 
27.89 & 172.24 & 1.0  & 132.54   & 172.6  & 0.83 \\
\rowcolor[HTML]{E4D3D2} 
27.89 & 54.47  & 10.0 & 235.69   & 122.75 & 0.91 \\
\rowcolor[HTML]{E4D3D2} 
27.89 & 31.45  & 30.0 & 310.18   & 108.4  & 0.93 \\
\hline
\rowcolor[HTML]{E4D3D2} 
33.52 & 248.65 & 1.0  & 91.81    & 228.13 & 0.72 \\
\rowcolor[HTML]{E4D3D2} 
33.52 & 78.63  & 10.0 & 163.26   & 157.46 & 0.86 \\
\rowcolor[HTML]{E4D3D2} 
33.52 & 45.4   & 30.0 & 214.86   & 136.39 & 0.89 \\
\hline
\rowcolor[HTML]{E4D3D2} 
39.14 & 339.05 & 1.0  & 67.33    & 290.25 & 0.59 \\
\rowcolor[HTML]{E4D3D2} 
39.14 & 107.22 & 10.0 & 119.73   & 196.23 & 0.78 \\
\rowcolor[HTML]{E4D3D2} 
39.14 & 61.9   & 30.0 & 157.57   & 167.59 & 0.84 \\
\hline
\rowcolor[HTML]{E4D3D2} 
56.0  & 694.18 & 1.0  & 32.89    & 512.99 & 0.19 \\
\rowcolor[HTML]{E4D3D2} 
56.0  & 219.52 & 10.0 & 58.48    & 334.73 & 0.49 \\
\rowcolor[HTML]{E4D3D2} 
56.0  & 126.74 & 30.0 & 76.96    & 278.54 & 0.61 \\
\hline
\end{tabular}
 \end{center}
\end{table}

Table \ref{tab:final_ao_calc} describes the engineering and AO residuals for examples of note and highlights a handful of optimized designs; one median example is an actuator spacing of $d_a=13.84$cm, mirror facesheet thickness of $t=13.41$mm, resulting in a quilting error of $\sigma_Q\leq10$nm, a final high order wavefront error of $\sigma_{WFE}\leq70.36$nm, and Strehl in the IR of $0.97$. Figure \ref{fig:strehls_wvs} shows Strehl for this system at additional science wavelengths. 

\section{CONCLUSIONS AND science applications}

In conclusion, we find feasible designs for adaptive primary telescope systems for mid-sized telescopes constructed with HVR actuators. In particular, using an actuator spacing of 14 cm provides workable designs, both in mirror facesheet thickness (7.9mm - 43.4mm) and the ability to correct for Mauna Kea-like seeing conditions (with a Strehl of 0.97). Such a device might prove scientifically productive in areas such as ground-layer AO correction, background-limited infrared observations, and visible adaptive optics correction performed with high actuator counts. In the following subsections we summarize potential science applications of note. 

\subsection{Ground-Layer Adaptive Optics Correction for Wide Field Surveys}

Wide-field adaptive optics correction can provide survey telescopes a means to improve the image quality of multiple targets simultaneously.  This approach requires sensing and correcting the portion of atmospheric turbulence corresponding to the ground-layer.  Limitations in this approach for post-focal plane AO systems arise with the complexity of the optical system needed to reimage a large ($>$10 arcmin) field-of-view.  Adaptive secondary mirrors solve this problem, but particularly on large telescopes, are conjugate to above or below the ground-layer, reducing their effective field-of-view for ground-layer AO (GLAO) correction.  Such a tradeoff can make a medium aperture telescope competitive with a larger one.  For example, a 2-m telescope that can correct fields of view up to 10-arcmin will obtain the same SNR faster than a 8-meter telescope limited to correcting 2-arcmin.\cite{Chun2014}.  
 


\subsection{Background-limited Infrared Imaging }

Each element in the optical path causes a small loss in throughput and adds thermal noise. Traditional adaptive optics systems that are located post-focal plane add five or more elements. 
For example, the 
Keck NIRC2 AO-system adds seven extra reflections to the optical path and accounts for approximately one-half of the thermal background noise present in mid-IR images (4.7um).  Although small, this noise and throughput loss can prove detrimental to sensitive photon-limited science, especially in the infrared. This can impact our ability to detect faint objects like exoplanets and brown dwarfs. 
The benefits of integrated AO systems to sensitive infrared science have been shown using adaptive secondary mirrors at The Large Binocular Telescope~\cite{Esposito2010}, VLT Observatory~\cite{Biasi2012}, and the Magellan Telescope~\cite{Close2018}. 

Using an adaptive primary mirror  
 provides a method to integrate the AO system into the telescope itself.  This would streamline the optical path and reducing the thermal background and throughput loss, similar to the impact of using an adaptive secondary mirrors.

\subsection{High-density actuator spacing AO systems}

In this paper, we determine the feasibility of building adaptive primary mirrors with the minimum number of actuators. However, using tight physical actuator spacing (spacing of 4-10cm) and increasing the number of actuators may benefit AO applications with needs for high-actuator counts.  This includes visible light AO, which requires a high actuator stroke and faster AO correction due to its shorter wavelengths.\cite{Hardy1998} 
Visible light AO systems have been in growing demand since 2013 and will be the only option for visible diffraction-limited observing after Hubble retires.~\cite{Close2016} 
Major observatories are investing in instruments that can perform these corrections, including but not limited to, Subaru’s SCExAO\cite{Jovanovic2016}, the MagAO-2k upgrade \cite{Males2016}, and LBT’s SHARK-VIS.\cite{Farinato2016} 

This trend towards commissioning high-density actuator spacing AO systems is expected to continue in ELT-class (and larger) telescopes.  However,  
creating an AO system with high actuator density becomes increasingly difficult as the primary mirror size of the telescope grows and typical post-focal plane AO systems (or even secondary mirrors) lack enough space to pack actuators.  Moving the AO system's deformable mirror to the primary mirror provides additional area to space the actuators. Additionally, large telescopes under development are likely to incorporate active optics; an adaptive primary mirror could provide a two-in-one solution for both the active optics and adaptive optics corrections.      
This idea was previously introduced  
for extremely large telescopes~\cite{Riccardi2004}, however, the technique was not adopted.
 While an adaptive primary could be feasible for future ELTs, it is difficult for major facilities to consider the option seriously before an adaptive primary is proven on sky. A mid-sized telescope equipped with an adaptive primary mirror could provide the first step towards a technological demonstration of adaptive primary mirrors needed for ELTs and beyond.

\appendix    

\section{AO Error Residual Calculations}
Our formulation of AO error residuals is derived from an John Hardy's ``Adaptive Optics for Astronomical Telescopes"\cite{Hardy1998}, as well as an adaptation made by Don Gavel and Katie Morzinski for the Adaptive Optics Summer School held at UC Santa Cruz.  


\label{sec:ao_appendix}

The overall Strehl ratio is calculated by, 
\begin{equation}
    \mathrm{Strehl} = \exp(- 2\pi {\sigma_{h.o.w}} ^2/\lambda  )
\end{equation}

\noindent where $\lambda$ is the science wavelength and $\sigma_{h.o.w}$ is the high order wavefront error.  For much of this study, we use $\lambda = 2.5\mu m$ (K-band). The high order wavefront error is a combination of four errors added in quadrature, 

\begin{equation}
    \sigma_{h.o.w} = \sqrt{\sigma_{fit}^2+\sigma_{meas}^2+\sigma_{BW}^2 + \sigma_{Q}^2}
\end{equation}

\noindent where $\sigma_{fit}$ is the fitting error, $\sigma_{meas}$ is the measurement error, $\sigma_{BW}$ is the bandwidth error, and $\sigma_{Q}$ is the quilting error.  We do not consider the case of laser guide stars or off-axis natural guide stars in this analysis. We ignore any affects of cone effect error (which would be present with laser guide stars) and anisoplanetic angle (present for off-axis guide stars). 
We discuss how we quantified the fitting, measurement, and bandwidth errors in the subsequent sections. (Quilting error is discussed previously in Section \ref{sec:eng_outline}.)

\subsection{Fitting Error} 
The fitting error ($\sigma_{fit}$) is calculated by

\begin{equation}
\sigma_{fit}=\frac{\lambda_0}{2\pi} \sqrt{\mu ({d_a}/{r_0})^{5/3} }  
\end{equation}

\noindent where $d_a$ is the actuator spacing projected on the primary, $r_0 = 15cm$ is the turbulence coherence cell size, and $\mu$ is the fitting parameter.  We use a fitting parameter value of $\mu = 0.3$, corresponding to a continuous deformable mirror geometry as quoted in Hardy 1998.~\cite{Hardy1998}.  The seeing conditions (including $r_0$) are defined with respect to a chosen wavelength $\lambda_0$, which we defined at is $\lambda_0$ = 500nm.

\subsection{Measurement Error}

The measurement error ($\sigma_{meas} $) is calculated by,  

\begin{equation}
    \sigma_{meas} =  \eta \chi d_{a} \frac{\sigma_{spot}}{SNR}
\end{equation}

\noindent where  $\eta$ is the noise propagator, $\chi = 0.61$ is the closed loop averaging factor, $d_{a}$ is the actuator spacing, and $\frac{\sigma_{spot}}{SNR}$ is the centroiding error.

The noise propagator ($\eta$) is calculated for square subaperture sizes with the equation from Noll 1978~\cite{Noll1978}, 

\begin{equation}
    \eta = \sqrt{2 (0.0536+0.0795\ln(N_{DOF}))}
\end{equation}

\noindent where $N_{DOF}$ is the reconstructors degrees of freedom in units of subapatures: $N_{DOF} = round(\frac{\pi}{4} (\frac{D}{d_a})^2)$.\cite{Noll1978} $D$ is the diameter of the telescope primary. 

We assume a natural guide star when calculating the hartmann spot size ($\sigma_{spot}$),  
\begin{equation}
    \sigma_{spot} = max(\frac{206265*(5.89*10^{-7})}{d_a}, 
     \frac{206265*(6.5*10^{-7})}{r_0})
\end{equation}

We calculate the signal-to-noise (SNR) by 
\begin{equation}
    SNR = \frac{N_{photons}}{\sqrt{N_{photons} +N_{pix}\sigma_{read}^2}}
\end{equation}

\noindent where the read noise is $\sigma_{read}$ = 1 electron/pixel, the number of pixels per subapature is $N_{pix} = 4$ pixels, and $N_{photons}$ is the number of photons collected on the AO guide camera in units of photons/frame/subapature.    The number of photons ($N_{photons}$) was quantified by 

\begin{equation}
    N_{photons} = B  (t_{samp})  (d_a)^2
\end{equation}

\noindent where the sample time $t_{samp}[s]= \frac{1}{10 f_c}$, $d_a$ is the actuator spacing, and the brightness of the guide star is $ B = (8.9*10^5)  10^{(0-m_V)/2.5}$ with a guide star magnitude of $m_V = 8$.






\subsection{Bandwidth Error} 

The bandwidth error is dependant on the control frequency ($f_c$) and the greenwood frequency ($f_g$), 

\begin{equation}
    \sigma_{BW} = \frac{\lambda_0}{2\pi}\sqrt{ \left( \frac{f_g}{f_c}\right)^{5/3}}
\end{equation}

\noindent We calculate the greenwood frequency by,
\begin{equation}
    f_g  = 0.426 \frac{v}{r_0(\zeta)}
\end{equation}

\noindent where $v = 10$m/s is the mean wind speed and $ r_0(\zeta) = r_0[\cos{(\zeta \frac{\pi}{180})}]^{3/5}$ is the value of our chosen $r_0$ (15cm) at our chosen zenith angle ($\zeta = 20^{\circ}$).  Our $r_0$ value is defined at a wavelength of $\lambda_0$ = 500nm. 

In cases where the resonant frequency ($f_r$) of the system is less than 10 times the control frequency of 50Hz, we replace the control frequency with $f_c = {f_r}/{10}$ to calculate how decreasing the resonant frequency impacts the bandwidth error.

\section{WIGGLE}
\label{sec:wiggle_appendix}

\begin{figure}[h!]
   \begin{center}
   \begin{tabular}{c}
  \includegraphics[width=\textwidth]{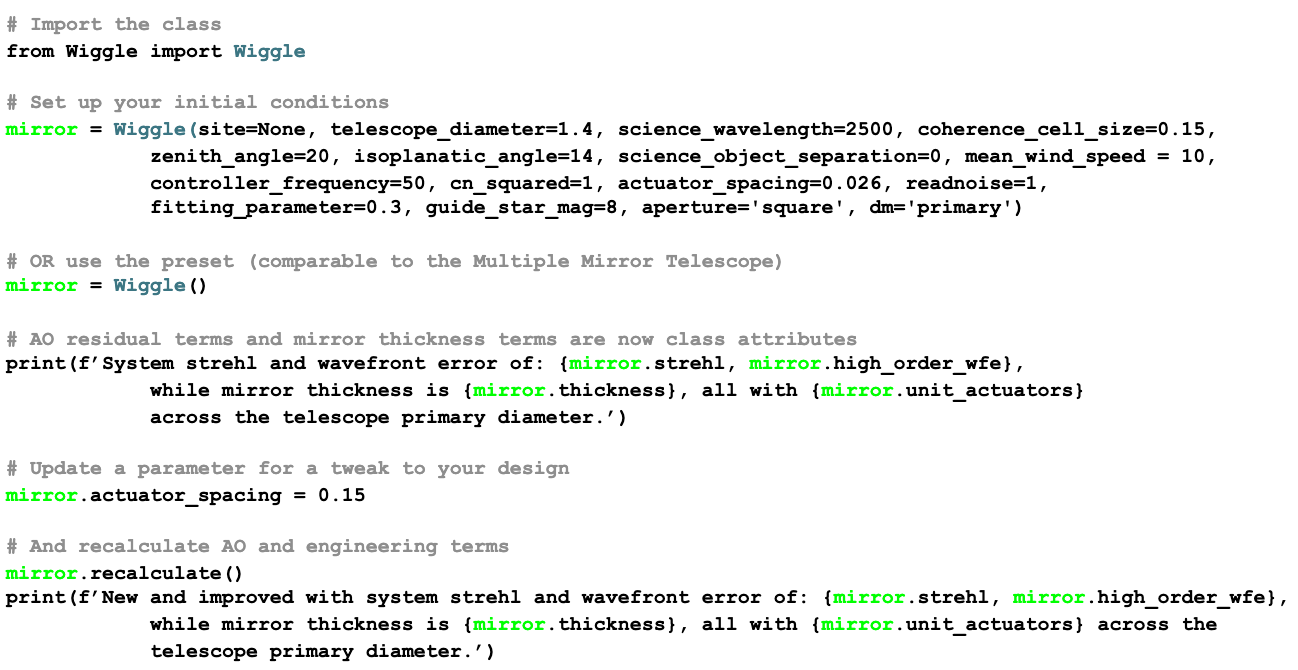}
   \end{tabular}
   \end{center}
   \caption[example] 
   { \label{fig:strehls_wvs} 
   \textbf{\texttt{WIGGLE} in practice.} This snippet of code shows an example of of setting up a mirror and calculating the AO residuals and engineering terms for an adaptive primary.}
   \end{figure}

\texttt{WIGGLE}, the Wavefront Improvements with GiGantic Light-deforming Experiments package is an open source pure-Python package to encapsulate AO performance calculations, available at \href{https://github.com/julesfowler/wiggle}{github.com/julesfowler/wiggle}. \texttt{WIGGLE} is based quite closely on math outlined in ``Adaptive Optics for Astronomical Telescopes"\cite{Hardy1998}, and informed by AO budget experiments designed for the annual Center for Adaptive Optics' AO Summer School.

\acknowledgments 
Thanks to the many folks who donated their time to talk through this concept, including Becky Jensen-Clem and Maaike van Kooten. Thanks to Mark Chun for his interesting suggestions on science cases for an adaptive primary system. Thanks to Katie Morzinski and Don Gavel for putting together and maintaining AO error budgets for the Center for Adaptive Optics Summer School, which heavily influenced this work.

\bibliography{report} 
\bibliographystyle{spiebib} 

\end{document}